\documentclass[%
APS,
twocolumn,
reprint,
superscriptaddress,
 amsmath,amssymb,
 aps,
]{revtex4}
\usepackage{amssymb,amsmath}
\usepackage{epsfig}
\usepackage{graphics}
\usepackage{graphicx}
\usepackage{dcolumn}
\usepackage{bm}
\usepackage{epstopdf}

\begin{document}


\title{Largely enhanced photogalvanic effects in the phosphorene photodetector by strain-increased device asymmetry}

\author{Juan Zhao}
\affiliation{Department of Physics, Shanghai Normal University, Shanghai 200234, China}
\author{Yibin Hu}
\affiliation{ Shanghai Institute of Technical Physics, Chinese Academy of Sciences, China}
\author{Yiqun Xie}
\email{yqxie@shnu.edu.cn}
\affiliation{Department of Physics, Shanghai Normal University, Shanghai 200234, China}
\author{Lei Zhang}
\affiliation{State Key Laboratory of Quantum Optics and Quantum Optics Devices, Institute of Laser Spectroscopy, and Collaborative Innovation Center of Extreme Optics, Shanxi University, Taiyuan 030006, China}
\author{Yin Wang}
\affiliation{Department of Physics and International Centre for Quantum and Molecular Structures, Shanghai University, 99 Shangda Road, Shanghai 200444, China}
\date{\today}

\begin{abstract}
 Photogalvanic effect (PGE) occurring  in  noncentrosymmetric materials  enables the generation of the open-circuit voltage that is much larger than the bandgap, making it rather attractive in  solar cells. However, the magnitude of the PGE photocurrent is usually small, which severely  hampers its practical application. Here we propose a mechanism to largely enhance the PGE photocurrent by mechanical strain based on the quantum transport simulations for the two-dimensional nickel-phosphorene-nickel photodetector.  Broadband PGE photocurrent governed by the $C_s$ noncentrosymmetry is generated at zero bias  under the illumination of linearly polarized light. The photocurrent depends linearly  on the device asymmetry, while nonlinearly on the optical absorption.  By applying the appropriate mechanical  tension stress on the phosphorene, the photocurrent can be substantially enhanced by up to 3 orders of magnitude,  which is primarily ascribed to the largely increased device asymmetry.  The change in the optical absorption in some cases can also play a critical role in tuning the photocurrent due to the nonlinear dependence.  Moreover, the photocurrent can even be further enhanced by the mechanical bending, mainly owing to the considerably enhanced device asymmetry.  Our results reveal the dependence of the PGE photocurrent  on the device asymmetry and absorption in transport process through a device, and also explore the  potentials of the PGE in  the  self-powered  low-dimensional flexible optoelectronics, and low-dimensional photodetections with high photoresponsivity.
\end{abstract}

\maketitle
\section{\textbf{INTRODUCTION}}
Optoelectronic properties of  the two dimensional (2D) materials have  attracted intensive research interest in the past decades for their fascinating potential applications in various fields including photodetections,\cite{AFM2019-29-1803807,AM201830-1801164,APR2019,APR2019b,ACSNano.2019.13.9907}
solar cells\cite{Nat.2019,Science.2018.360} and spintronics.\cite{NC2019-10-3965,PNAS2016-113-3746,PRA2018}
There are several light-to-current conversion mechanisms that are usually involved in these 2D optoelectronic devices. The most typical mechanism  is the photovoltaic effect which requires the  traditional $p$-$n$ junctions to separate the light induced electron-holes. Their efficiency has been improved largely in these years and almost reaches the Shockley-Queisser limitation.\cite{Shockley} The second one is the photoconductor effect, in which the external voltage is necessary for generating  the persistent current, yet can not provide the electrical power generation.\cite{huang2016highly,furchi2014mechanisms} Other mechanisms like photo-gating effects are also involved.\cite{wang2014inkjet,qian2017high,cao2019enhanced}

In contrast to these light-to-current conversion mechanisms, the photogalvanic effect (PGE),\cite{Sov.Phys.Usp.23.199,APL.2002.77.3146,PgeJPCM.2003.15.20,nanoscale,xiePRA,ACS2020} also known as the bulk photovoltaic effect (BPVE),\cite{PRB.23.5590,APL,Sci.Adv.2019.5} enables the generation of the photocurrent without the need of any external voltage or the $p$-$n$ junctions, which occurs in materials with broken inversion symmetry. The PGE has attracted a broad researching interest in wide material systems, including the bulk Weyl semimetals,\cite{Nat.Mater.2019.18.9,Nat.Mater.2019.18.5} quantum wells,\cite{APL.2002.77.3146} and 2D materials.\cite{nat.Commun.2018.9}  This mechanism is able to generate an open-circuit voltage that is much larger than the bandgap,\cite{APL,Nat.Commun.8.281-2017} which  makes it possible to achieve the light-to-current efficiency exceeding the Shockley-Queisser limit.\cite{Nat.Photonics.2016.10.9}   However, high efficiency has been hardly achieved so far in experiments mainly because of the low magnitude of the PGE  photocurrent (photoresponsivity) which is usually less than mA/W. It is therefore rather desirable to find an effective approach to enhance the PGE photocurrent.

Breakthroughs  have been achieved recently. It has been demonstrated that the PGE photocurrent obtained in the 1D WS$_2$ nanotubes is several orders of magnitude larger than that in the flat 2D WS$_2$ mono-layers,\cite{Nat.2019} and the photoresponsivity is of up to 0.1 A/W, which is much larger than those PGE ever reported in other systems, such as the bulk BaTiO$_3$\cite{Nat.Photonics.2016.10.9}, TTF-CA\cite{dressel2017infrared} and organometallic perovskite halide\cite{APL.2016.109}. This substantially enhanced photocurrent in the 1D WS$_2$ nanotube  has been mainly ascribed to its reduced $C_{2nv}$ or $C_{2nh}$ symmetry, as compared to the 2D WS$_2$ monolayer ($D_{2h}$), along with the increased optical absorption. These results suggested that reducing the crystal symmetry of a flat 2D semiconducting material by creating a curvature ({\it e.g.}, bending) can strengthen the PGE and hence enhance the photocurrent. Another important experiment has demonstrated that by applying appropriate mechanical stress gradient on the SrTiO$_3$ single crystal the PGE photocurrent can  be enhanced by orders of magnitude,\cite{Science.2018.360} which is known as the flexo-photovoltaic effect. However, controversy has been raised regarding the  fundamental mechanism responsible for this enhancement in photocurrent.\cite{ACSNano.2019.13.12259} A direct experimental evidence for the effective enhancement of the PGE by mechanical strain has been achieved recently in the Fe-doped LiNbO3 single crystal, in which the photocurrent was increased by 75\% even under a tiny uniaxial compressive stress (0.005\%) on the lattice.\cite{sci.adv.2019.5.eaau9199} Unfortunately, much higher compressive stress is difficult to be achieved experimentally so that the further enhancement of the photocurrent has not been investigated. Therefore, it remains an open question at the present that whether the PGE photocurrent can be increased by orders of magnitude under appropriate mechanical stress or bending due to the lack of the direct experimental supports.

Theoretical studies have revealed  that the PGE involves two mechanisms, one of which is connected with the asymmetry in the electron velocity distribution.\cite{Sov.Phys.Usp.23.199,BookPGE,Jin,Wangbin} Another mechanism stems from the displacement of the electrons in real space when they undergo quantum transitions, namely, shift current.\cite{BookPGE,shift1982,Nat.Commun.8.281-2017,PgePRL.2017.119.6,PRL2012shift1,PRL2012shift2,SciAdv2016shift}
These excellent theoretical achievements on the PGE have achieved much success in understanding experimental results, and provided deep insights into the physical origin of thePGE.\cite{PgePRB.2016.93.8,PgePRB.2017.96.7,PgePRB.2019.99.7,PgePRB.2019.100.6,PgePRL.2017.119.6,PgePRL.2019.122.16,PgePRL.2020.124.7.196603,PgePRL.2020.124.7.077401,PgePRX.2014.4.3} However, these studies focus mainly on the PGE in periodical systems (crystals), while in practical applications the PGE photocurrent transports  through the device which is an open system, and therefore involves statistical and nonequilibrium  physics that beyond the calculations for a periodical system. Therefore, how to enhance the PGE photocurrent flowing through the device under mechanical strain deserves further theoretical investigations.

In this study, we proposed a mechanism to enhance the PGE photocurrent generated in the device by quantum transport calculations based on  the 2D phosphorene photodetector with the nickel electrodes under the vertical illumination of linearly polarized light. A parameter for the device asymmetry was presented, which plays an important role in determining  the strength of the PGE. The photocurrent can be substantially enhanced by 3 orders of magnitude  by applying the appropriate mechanical tension stress and bending, owing to the largely increased device asymmetry.
\section{Model and methods}
\begin{figure}[htbp]
	\begin{center}
		\includegraphics[width=9cm]{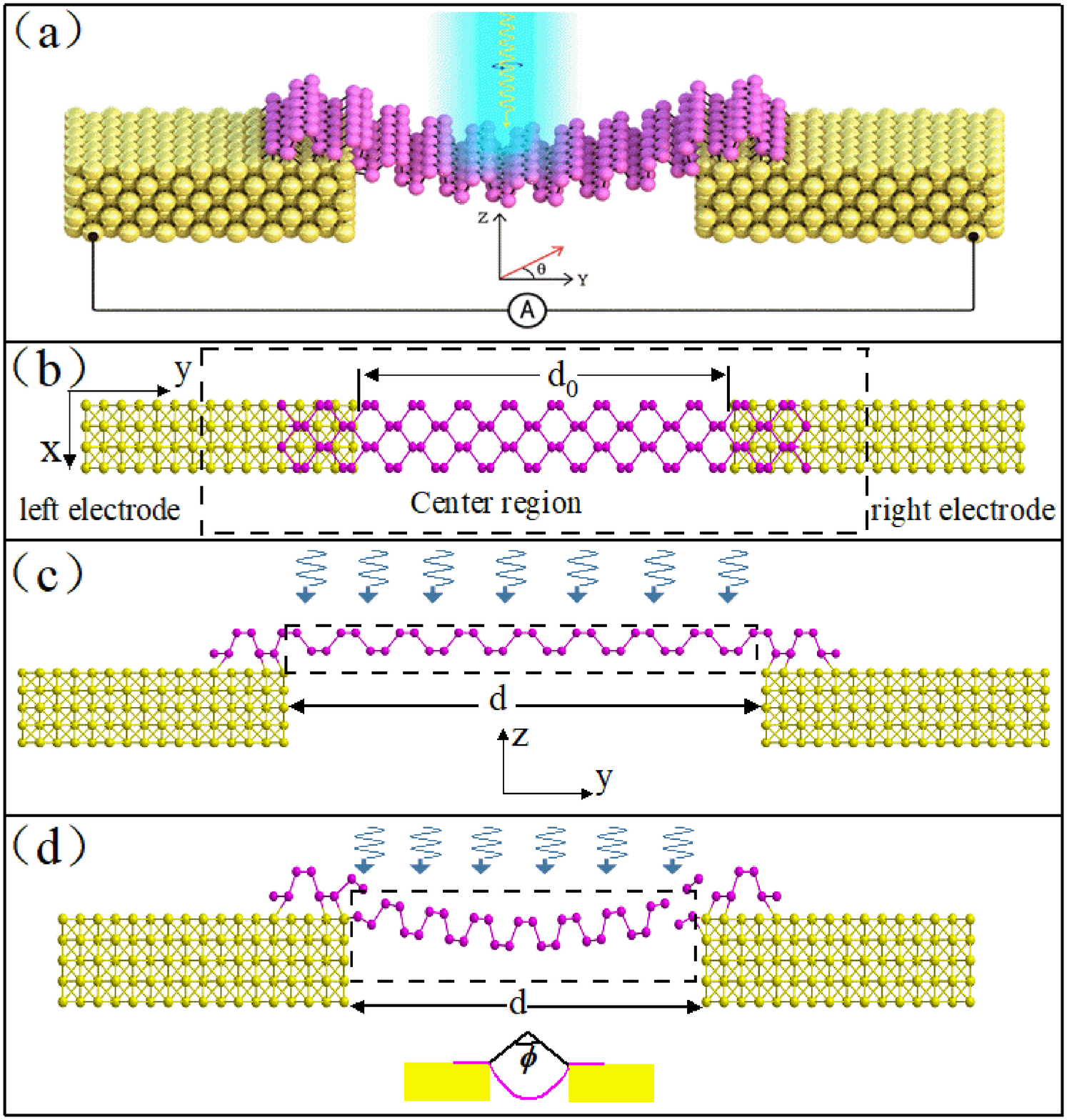}
	\end{center}
	\caption{ (a) Illustration of the  Ni(100)-phosphorene-Ni(100) photodetector under the mechanical bending irradiated by the polarized light. (b) The top view of the device model for the Ni(100)-phosphorene-Ni(100) photodetector without any strains, and (c) the side views for the photodetector with the tension stress, and that (d) with the bending ratio of 3.9\%, which corresponds to a central angle $\phi=60^{\circ}$, as indicated in the inset. The  $d$ and $d_{0}$ denote the distance between the two leads with and without mechanical deformations, respectively. Yellow and pink spheres denote the Ni atoms and P atoms, respectively. The gray wavy lines with arrows denote the incident light irradiated on the free standing phosphore indicated by the shadows.}
	\label{fig1}
\end{figure}

We modeled  a photodetector device with the Ni(100)-phosphorene-Ni(100) configuration, which contains the right/left electrode and the center region, as shown in Figs.~\ref{fig1}(a,b).
In the center region of the device, the phosphorene is partially overlapped with the Ni(100) electrodes which extend to $y=\pm\infty$, as shown in Fig.~\ref{fig1}(b). The lattice constants are $a_x=3.524 $ \AA\ and $a_y=4.580 $ \AA\ for the phosphorene calculated using the VASP code.\cite{vasp,DFT}  To build a periodic structure the phosphorene is uniformly stretched by about 6\% along the $x$ direction to match the Ni lattice. The Ni electrodes are modeled by a five-layer Ni(100) slabs, and the spins of the two ferromagnetic Ni electrodes are set in a parallel configuration. The distance between the phosphorene and the Ni slab surface is 1.95 {\AA}. The phosphorene-Ni(100) contact breaks the intrinsic spacing inversion symmetry of the phosphorene, as discussed in our previous work.\cite{xie-two.Phys.Rev.Appl.2018.10} The whole device structure  belongs to  the noncentrosymmetric $C_s$ group, which has only one mirror reflection plane located in the $y$-$z$ plane, considering the periodicity in the $x$ direction. Therefore, the PGE can be induced in the photodetector to generate the photocurrent flowing along the $y$ direction  when the free-standing phosphorene in the center region is vertically irradiated by linearly polarized light.

It has been shown that the monolayer phosphorene has an outstanding flexibility under both the in-plane and out-of-plane mechanical strain.\cite{APL.2014104.25,NanoLett.2015.15.3,Nanotechnology.2016.27.5} In our simulations, the mechanic tension is applied uniformly on the phosphorene along the $y$ direction by increasing the $y$-distance $d$ between the two electrodes, and the mechanical bending is applied along the out-of-plane direction with a certain central angle $\phi$. When the  mechanical strain is applied, the distance between the two electrode will change. Accordingly, we can define the ratio for the applied tension stress ($\varepsilon_{y}$) and bending ($\varepsilon_{b}$)  as the $\varepsilon_{y,b}$=$(d-d_0)/d_0\times100\%$, where $d_0$ = 37.77{\AA} when no strain is applied [See Fig.~\ref{fig1}(b)]. For example, for the bending with the central angle $\phi$ of 60$^\circ$, as illustrated in the inset of Fig.~\ref{fig1}(d),  the  bending ratio  is 3.9\%. The deformed phosphorene was fully relaxed by VASP with the atoms in the Ni electrodes being fixed. In our VASP calculations, the exchange-correlation energy was treated by the projector augmented wave of the Perdew-Burke-Ernzerhof\cite{PBE} with an energy cutoff of 500 eV. The Brillouin zone was sampled with a 8$\times$1$\times$1 mesh of the Monkhorst-Pack k-points.\cite{Mo} A vacuum region of 20 \AA\ in the $z$ direction is added in the 2D supercell to isolate any possible spurious interaction between periodical images of the supercell.

When the center region is irradiated by the polarized light,  the photocurrent can be generated. Based on the linear response approximation,\cite{JAP2002} the photocurrent injecting into the left electrode can be written as,\cite{chenjz,xieS,ZhangL2}
\begin{equation}
J_L^{(ph)} = {ie\over h} \int \textrm{Tr}\{\Gamma_L[G^{<(ph)}+f_L(E)(G^{>(ph)}-G^{<(ph)})]\}\textrm{d}E, \label{I}
\end{equation}
where $G^{>/<(ph)}(E)$ is the greater/lesser Green's function including electron-photon interactions, and $f_L$ is the Fermi-Dirac distribution function.  To analyse the PGE,  we can assume that the temperature is low enough and hence the Fermi function $f_L$ is stepwise, so that the Eqn. (1) can be rewritten as,\cite{zhangL}
\begin{eqnarray}\label{I}
J_L^{(ph)}& =& {ie\over h} \int^0_{-\infty} \textrm{Tr}[\Gamma_L G^{>(ph)}]\textrm{d}E+ \int_0^{+\infty} \textrm{Tr}[\Gamma_L G^{<(ph)}]\textrm{d}E \nonumber\\
&= &J_-+J_+,
\end{eqnarray}
where $J_{-}= \int^0_{-\infty} \textrm{Tr}[\Gamma_L G^{>(ph)}]\textrm{d}E <0$, denotes  the current flowing out the center region to  the left lead, while $J_{+}= \int_0^{+\infty} \textrm{Tr}[\Gamma_L G^{<(ph)}]$ $ \textrm{d}E$ $>0$ denotes the current flowing into the center region from the lead. The magnitude of the PGE photocurrent is thus determined by the $|J_L^{(ph)}|=\Delta J=|J_{+} -|J_{-}||$.  Therefore we can define a factor $A$ to describe the device asymmetry as
\begin{equation}
\textit{A} =  \left| \frac{J_{+}-|J_{-}|}{J_{+}} \right|. \label{I}
\end{equation}

For the device with the space inversion symmetry the current flowing into  the center region equals that out of the center region, that is, there is no net current flowing through the device. In this case,  $J_{+}=|J_{-}|$, so that $J_L^{(ph)}=0$, and  $A$=0. For the device without the inversion symmetry, there is a net photocurrent and  $A$ is nonzero. The magnitude of the photocurrent can then be rewritten as,
\begin{equation}
|J_L^{(ph)}|= AJ_{+}. \label{J}
\end{equation}
Note that the $J_{+}$ includes the $\Gamma_L G^{<(ph)}$, while the $ G^{<(ph)}$
includes the photon propagator $D^{<}$,\cite{JAP2002}
\begin{equation}
 D_{ln;qm}^<(E)=2\pi M_{ln}M_{qm}N\delta(E-\hbar\omega),\label{D}
\end{equation}
where
\begin{equation}
 M_{ln}=\frac{e}{m}\left( \frac{\hbar\sqrt{\tilde{\mu_r}\tilde{\epsilon_r}}I_\omega}{2N\omega\tilde{\epsilon} c } \right)^{1/2} \left<l|\textbf{p}\cdot\textbf{e}|n\right>.\label{M}
\end{equation}
Here, $l,n,m,q$ are the labels of the atomic orbital, $I_\omega$ is the photo-flux,  $m_0$ is the bare electron mass, $e$ the electronic charge, $\tilde{\mu_r}$  the relative magnetic susceptibility, $\tilde{\epsilon_r}$ the relative dielectric constant, and $\tilde{\epsilon}$  the static dielectric constant, $\textbf{p}$ the electron momentum, and \textbf{e} the light polarization vector. Therefore, in essentials the  photon propagator  $D^{<}$  corresponds to the excitation of  an electron by the absorption of a photon.\cite{JAP2002,Datta1992} As a comparison, we give the optical absorption coefficient $\alpha$ of a material,\cite{PRB.23.5590,book}
\begin{eqnarray}\label{alpha}
 \alpha(\omega)&=&\frac{e^2}{2\pi^2\epsilon_0 m_0^2\omega c}\sum_{n,l}\int(f_n-f_l)|\left<\textbf{k},l|\textbf{p}\cdot\textbf{e}|\textbf{k},n\right>|^2 \nonumber\\
 &\times&\delta(E_l-E_n-\hbar\omega)\textrm{d}^3k.
\end{eqnarray}
It can be seen that the photon propagator $D^{<}$ and the optical absorption coefficient $\alpha$  have the same physical meaning, that is, the
excitation of electrons by absorbing the photon energy of $\hbar\omega$. This indicates that the  $J_{+}$ should have a nonlinear dependence on the optical absorption of the device due to the tensor product of the  $D^{<}$ with the coupling matrix $\Gamma$. Consequently,  the photocurrent $J_L^{(ph)}$ has a linear dependence on the device asymmetry $A$ (a scalar), while a nonlinear dependence on the optical absorption of the device,  according to  Eqn.~(\ref{J}).

The PGE photocurrent also varies with the light polarization. In our previous work, we have shown that for linearly polarized light the photocurrent has a dependence on the polarization angle $\theta$ and can be written as,\cite{xieS,xiePRA,xieJMCA}
\begin{eqnarray}\label{LPGE1}
J_L^{(ph)}&=&{ie\over h}\int\{\cos^2\theta\textrm{Tr}\{\Gamma_L[G^{<(ph)}_{1}+f_L(G^{>(ph)}_{1}-G^{<(ph)}_{1})]\}  \nonumber\\
&+&\sin^2\theta\textrm{Tr}\{\Gamma_L[G^{<(ph)}_{2}+f_L(G^{>(ph)}_{2}-G^{<(ph)}_{2})]\} \nonumber\\
&+&{2\sin(2\theta)}\textrm{Tr}\{\Gamma_L[G^{<(ph)}_{3}+f_L(G^{>(ph)}_{3}-G^{<(ph)}_{3})]\}\}\textrm{d}E, \nonumber
\end{eqnarray}

where $G^{>/<(ph)}_{1,2,3}$ depends on the photon frequency $\omega$,  the electron momentum $\textbf{p}$, and the  polarization vector \textbf{e}=$\cos\theta$\textbf{e}$_1$+$\sin\theta$\textbf{e}$_2$. In our calculations, the unit vectors $\textbf{e}_1$ and $\textbf{e}_2$ are set along the $x$ (zigzag) and $y$ (armchair) directions, respectively. The photocurrent can be further normalized as $J= J_L^{(ph)}/eI_\omega$, where $I_\omega$ is the photo-flux. Note that the normalized photocurrent $J$  still has a dimension of area, i.e,. $a_0^2$/phonon, where $a_0$ is the Bohr radius.
To facilitate the analysis of the enhancement of the photocurrent, we define three factors, namely, $R_J$, $R_A$ and $R_\alpha$, which denote respectively the increase ratio of the photocurrent $J$, device asymmetry $A$ and the optical absorption coefficient $\alpha$ with respect to those when no strain is applied.

\section{Results and discussion}
\begin{figure}[hptb]
	\begin{center}
		\includegraphics[scale=0.28]{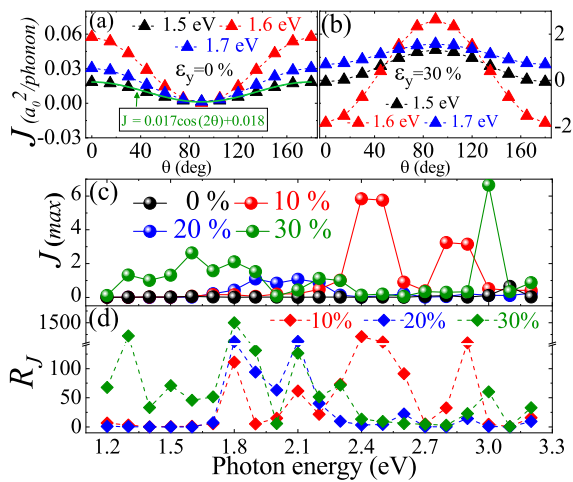}
	\end{center}
	\caption{The photocurrent $J$ for the photon energies of 1.5 eV, 1.6 eV and 1.7 eV without (a) and with (b) the mechanical tension, where the green solid line denotes the fitting function. (c) The variation of the maximum $J$ with the photon energy for different  tension ratio $\varepsilon_y$. (d) The increase ratio $R_J$ of the photocurrent  under the different mechanical tension with respect to that without the tensions. }
	\label{fig2}
\end{figure}

We calculated the PGE photocurrent  under different  mechanical stress for the Ni-phosphorene-Ni photodetector. The photon energy considered is from 1.2 eV to 3.2 eV, which is larger than the bandgap of the pristine phosphorene and covers the near infrared and visible range. We found that the photocurrent $J$ shows the cosine dependence on the polarization angle $\theta$ for all of the photon energies, which is a characteristic behavior of the PGE photocurrent governed by the $C_s$ symmetry.\cite{Sov.Phys.Usp.23.199}
Examples are presented in Fig.~\ref{fig2}(a)  for the photon energies of 1.5 eV, 1.6 eV and 1.7 eV, respectively, when no mechanical tension is applied. Specifically, the photocurrent for the photon energy of 1.5 eV is well fitted by the function of $0.017*\cos(2\theta)+0.018$, which agrees qualitatively with the phenomenological theory.\cite{Sov.Phys.Usp.23.199,APL.2002.77.3146}
We next examined the effect of the mechanical strain on the behavior of the photocurrent, and found that the photocurrent still preserves the cosine dependence on the $2\theta$ under the mechanical tension, since the mechanical tension applied uniformly along the $y$ direction does not change the $C_s$ symmetry.  Examples are shown in Fig.~\ref{fig2}(b) for the tension ratio of $\varepsilon_y=30\%$. However, it can be seen that the magnitude of the photocurrent is much larger than those without the mechanical tension. For comparisons, we obtained the maximum photocurrent  at $\theta=90^{\circ}$ or $ 0^{\circ}$ for all of the photon energies under the different tension ratios, as shown in Fig.~\ref{fig2}(c). It can be seen that  the magnitude of the photocurrent is substantially enhanced by the mechanical tension for most of the photon energy from 1.2 eV to 3.2 eV. More specifically, Fig.~\ref{fig2}(d) presents the increase ratio $R_{J}$ of the photocurrent under the mechanical tension with respect to that without the mechanical tension.  Noticeably, at the $\varepsilon_{y}=30\%$, the $R_{J}$   are all greater than 1 for almost all of the photon energy except 3.1 eV (see the green squares), and the $R_{J}$  exceed $10^2$ for a number of photon energies. Moreover,  the maximum ratio reaches to 1.5$\times10^3$ for the $\varepsilon_y=30\%$  at the photon energy of 1.8 eV. For the other $\varepsilon_{y}$,  the $R_{J}$  is also  greater than 1 for most of the photon energy.
\begin{figure}[htbp]
	\begin{center}
		\includegraphics[scale=0.28]{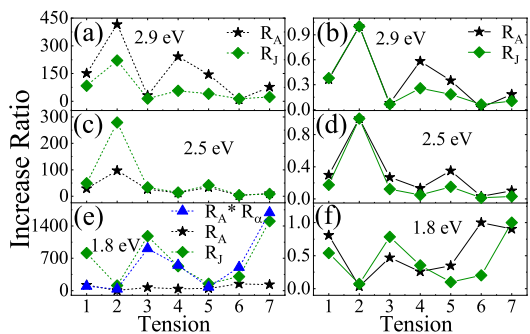}
	\end{center}
	\caption{ The increase ratios of the device asymmetry ($R_A$) and of the photocurrent ($R_J$) at the different  tension ratio $\varepsilon_y= (1)5\%, (2)10\%, (3)12.5\%, (4)15\%, (5)17.5\%, (6)20\%, (7)30\%$ for the photon energy of 2.9 eV (a), 2.5 eV (c) and 1.8 eV (e), respectively, and (b), (d), and (f) are the  normalized ratios. }
	\label{fig3}
\end{figure}

To understand this remarkable enhancement of the photocurrent under the mechanical tension, we first analyze  the device asymmetry $A$ and examine its behavior for  different tension ratio $\varepsilon_{y}$. The asymmetry $A$ is determined by the photon energy and also the tension ratio. We mainly focus on the photon energy of 2.9 eV, 2.5 eV and 1.8 eV, whose PGE photocurrent is enhanced most largely. The $A$ under different $\varepsilon_{y}$ and its increase ratio  $R_A$  are calculated for the $\varepsilon_{y}$ = $5\%, 10\%, 12.5\%, 15\%, 17.5\%, 20\%$ and $30\%$ with respect to that without the tension stress, respectively. Figs.~\ref{fig3}(a,c,e) show the variation of the $R_A$ (the black stars) and $R_J$ (the green squares) with the $\varepsilon_{y}$ for the photon energy of 2.9 eV, 2.5 eV and 1.8 eV, respectively, and Figs.~\ref{fig3}(b,d,f) give the normalized ratios with respect to their largest value. From the normalized ratios, we observed that the overall trend of the $R_J$ matches monotonously with that of the $R_A$ for the different $\varepsilon_{y}$. This indicates that the enhancement of the PGE photocurrent is primarily determined by the change in the device asymmetry $A$. Nevertheless, there are evident differences between the  $R_A$ and  $R_J$ at some $\varepsilon_{y}$, as shown in Figs.~\ref{fig3}(a,c,e).
It means that the enhancement of the photocurrent can not be exclusively attributed to the change in $A$. The influences from the change in the optical absorption induced by the mechanical tension should also be considered.  Therefore, we  calculated the  optical absorption coefficient $\alpha$  of the monolayer phosphorene and the increase ratio $R_\alpha$ under the different $\varepsilon_{y}$, as listed in Tab.I.

According to the Eqn.~(\ref{J}), in general the photocurrent should has a nonlinear dependence on the optical absorption. This nonlinear relationship is difficult to be expressed explicitly due to the tensor product of the $\Gamma$ and $G^{<ph}$, which varies with the different photon energy and the mechanical strain. For example, for the photon energy of 2.5 eV  (Fig.~\ref{fig3}(c)), the   change in the optical absorption seems to has a negligible  influence on the photocurrent since the  $R_A$ almost equals to the $R_J$  for most of the $\varepsilon_{y}$ except that at the $\varepsilon_{y}=10\% (2)$. In contrast, the $R_A$ is  evidently less than the $R_J$  for the photon energy of 1.8 eV  (Fig.~\ref{fig3}(e)), which indicates a different dependence on the optical absorption. Interestingly, we found that at most of the $\varepsilon_{y}$ the $R_J\approx R_A*R_{\alpha}$, as the blue triangles in Fig.~\ref{fig3}(e) shows. This indicates that for the photon energy of 2.5 eV, the photocurrent has a linear dependence on both the optical absorption and the asymmetry $A$ in most of the cases.

\begin{table*}[hpt]
	\small
		\caption{The optical absorption coefficient $\alpha$ and its increase ratio  $R_{\alpha}$ of the monolayer phosphorene under the  mechanical tension stress and bending. }
	\label{tbl:example2}
	\begin{center}
	\begin{tabular*}{1.0\textwidth}{@{\extracolsep{\fill}}llllllllll}
   		\hline
       Tension & $\varepsilon_{y}$ & 0$\%$ & 5$\%$ & 10$\%$ & 12.5$\%$ & 15$\%$ & 17.5$\%$ & 20$\%$ & 30$\%$ \\
              \hline
       1.8 eV & $\alpha$ & 0.344 & 0.274 & 2.286 & 4.536 & 5.034 & 0.510 & 1.177 & 4.348 \\
	          & R$_{\alpha}$ & & 0.798 & 6.648 & 13.191 & 14.639 & 1.483 & 3.424 & 12.644 \\
       2.5 eV & $\alpha$ & 1.521 &	1.892 &	3.651 &	5.073 & 5.558 & 4.046 & 6.410 & 6.246 \\
	          & R$_{\alpha}$ & & 1.243 & 2.400 &	3.334 &	3.652 &	2.659 &	4.213 &	4.105 \\
       2.9 eV & $\alpha$ & 4.585 &	4.142 &	5.469 &	3.267 &	1.927 &	2.077 &	8.390 &	5.248 \\
	          & R$_{\alpha}$ & & 0.903 & 1.193 &	0.713 &	0.420 &	0.453 &	1.830 &	1.145 \\

\hline
	       Bending  & $\varepsilon_{b}$ & 0$\%$ & 0.9$\%$ & 1.3$\%$ & 2.5$\%$ & 3.9$\%$ & 5.3$\%$ & 6.3$\%$ & 8$\%$ \\
\hline
     1.3 eV & $\alpha$ & 0.066 & 1.482 & 1.500 & 1.639 & 1.634 & 1.661 & 1.917 & 0.459 \\
	        & R$_{\alpha}$ & & 22.390 &	22.663 & 24.773 & 24.699 & 25.109 &	28.976 & 6.943 \\
     1.8 eV & $\alpha$ & 0.180 &	1.965 &	1.980 &	1.999 &	2.010 &	2.044 &	2.017 &	2.165 \\
	        & R$_{\alpha}$ & & 10.902 &	10.982 & 11.090 & 11.150 & 11.340 &	11.188 & 12.010 \\
     2.1 eV & $\alpha$ & 1.157 &	1.474 &	1.761 &	1.705 &	1.369 &	1.308 &	1.657 &	2.311 \\
	        & R$_{\alpha}$ & &1.273 &	1.522 &	1.473 &	1.183 &	1.130 &	1.432 &	1.996 \\
     \hline
	\end{tabular*}\label{tab2}
	\end{center}
\end{table*}

\begin{figure}[htbp]
	\begin{center}
		\includegraphics[scale=0.28]{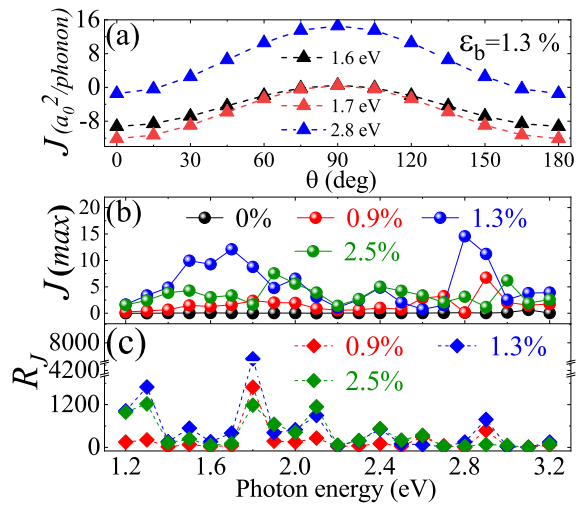}
	\end{center}
	\caption{(a) The photocurrent for the photon energies of 1.5 eV, 1.6 eV and 1.7 eV  under the mechanical bending. (b) The variation of the maximum photocurrent with the photon energy under the different bending ratio $\varepsilon_b$. (c) The increase ratio $R_J$ of the photocurrent under the different mechanical tension. }
	\label{fig4}
\end{figure}

We next consider the influence on the photocurrent by applying the mechanical bending on the free standing part of the phosphorene, as shown in Fig.~\ref{fig1}(d). Obviously, the space inversion symmetry of the phosphorene is broken by the bending, while the whole device still remains the $C_s$ symmetry. Note that the  mechanical tension does not change the symmetry of the phosphorenene. This means that the asymmetry of the device can be further increased by the mechanical bending, and therefore the PGE photocurrent would be enhanced further, as compared with that under the mechanical tension stress. We found that the photocurrent still holds the form of $\cos(2\theta)$ due to the unchanged  $C_s$ symmetry of the device,  as illustrated in Fig.~\ref{fig4}(a). Evidently, the magnitude of the photocurrent is largely increased, as compared with those without the mechanical bending, as shown in Fig.~\ref{fig4}(b). Moreover, the maximum photocurrent is 14.5 at 2.8 eV for $\varepsilon_b=1.3\%$, which is two times larger than that (6.6) under the tension stress (See Fig.~\ref{fig2}(c)).

\begin{figure}[htp]
	\begin{center}
		\includegraphics[scale=0.32]{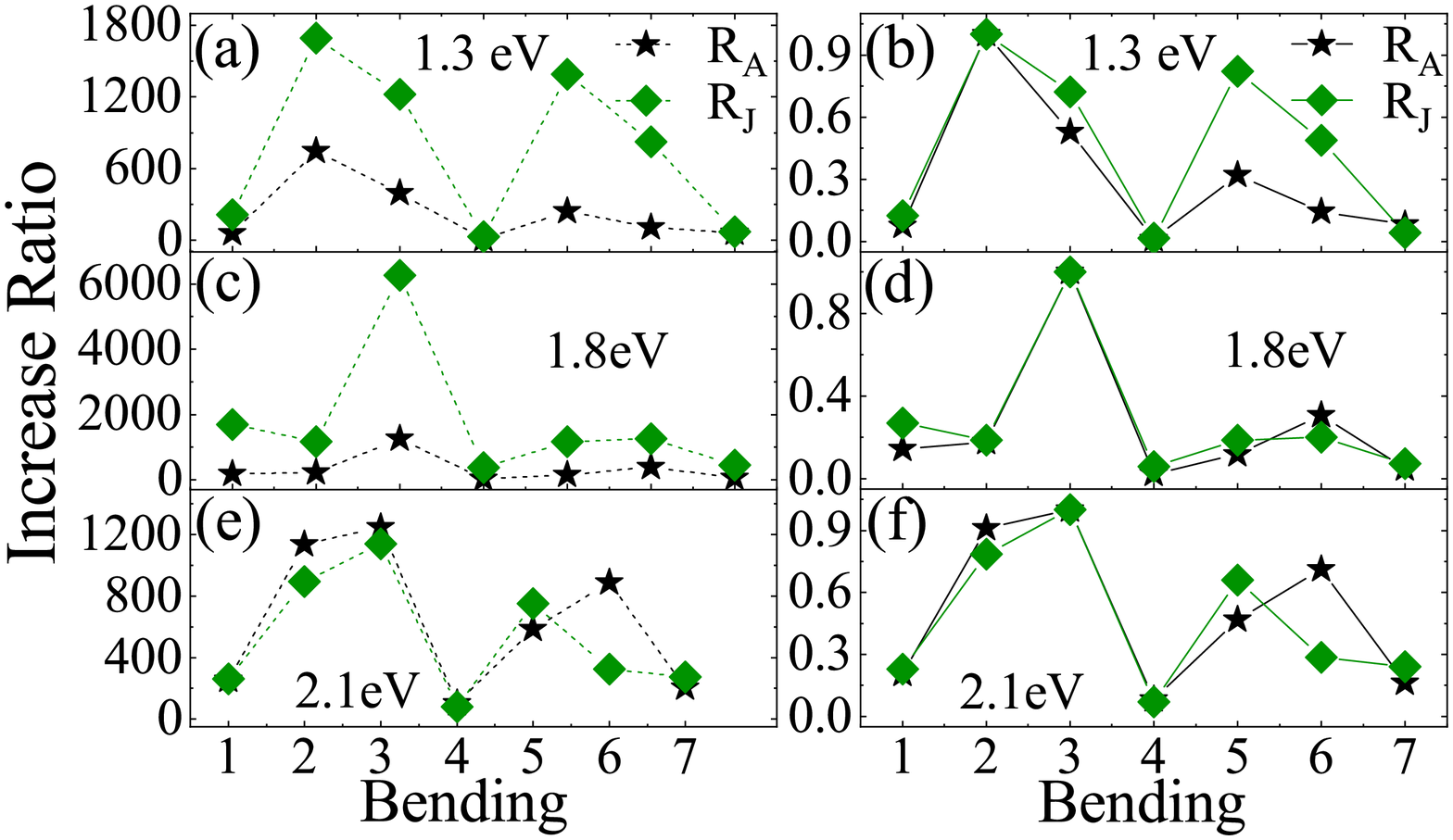}
	\end{center}
	\caption{ The increase ratios of the device asymmetry ($R_A$) and of the photocurrent ($R_J$) at the different bending ratio $\varepsilon_b= (1) 0.9\%, (2) 1.3\%, (3) 2.5\%, (4) 3.9\%, (5) 5.3\%, (6) 6.3\%, (7) 8\%$ for the photon energy of 1.3 eV (a), 1.8 eV (c) and 2.1 eV (e), respectively, and (b), (d), and (f) are the  normalized ratios.  }
	\label{fig5}
\end{figure}

We observed that for a number of   photon energies the  $R_J$ is greater than 10$^3$, and the maximum $R_J$  reaches about 6.26$\times10^3$ for the photon energy of  1.8 eV at the $\varepsilon_{b}=1.3\%$ (the blue squares), which is about 4 times larger than that under the mechanical tension. At the photon energy of 1.3 eV, 1.8 eV and 2.1 eV, the photocurrent show the most large enhancement.  We therefore give in Figs.~\ref{fig5}(a,c,e) the $R_A$ and $R_J$ for these photon energies at different mechanical bending ratio $\varepsilon_{b}$ of  (1) 0.9\%,  (2) 1.3\%, (3) 2.5\%, (4) 3.9\%, (5) 5.3\%, (6) 6.3\% and (7) 8\%, and correspondingly give in Figs.~\ref{fig5}(b,d,f) the normalized ratios. It can be seen that the trend of the $R_J$ is in a good agreement with that of the $R_A$. This indicates that the remarkable enhancement of the photocurrent should also  be mainly  attributed to the increased device asymmetry $A$. Nevertheless, the magnitude of the $R_A$ still differs from the $R_J$.  For the photon energy of 1.3 eV and 1.8 eV the $R_A$  is evidently less than the $R_J$  at the different $\varepsilon_{b}$, as shown in  Figs.~\ref{fig5}(a,c). Therefore, we should consider the influence of the change in the optical absorption.  We found that the optical absorptions are all  increased evidently  at the different $\varepsilon_{b}$ for the photon energy of 1.8 eV and 1.3 eV, as listed in Tab.I. This means that the photocurrent  has a positive dependence on the optical absorption. In contrast, for the photon energy of 2.1 eV the $R_A$ has a much less deviation  from the $R_J$ (Fig.~\ref{fig5}(c)), which indicates that the change in the optical absorption has a very small influence on the photocurrent. Accordingly, we found that $1<R_{\alpha}<2.0$ at every $\varepsilon_{b}$, which means that the optical absorption changes slightly under the mechanical bending for the photon energy of 2.1 eV. Besides, it is worth noting that  under the mechanical bending the largest $R_A$ is 1254 at the $\varepsilon_{b}$=2.5\% for the photon energy of 1.8 eV ( see Fig.~\ref{fig5}(c)), while the largest $R_A$ is 415 achieved under the mechanical tension stress of $\varepsilon_{y}=10\%$ for 2.9 eV (see Figs.~\ref{fig3}(a)). This means that the device asymmetry is  indeed further strengthened by the mechanical bending since it   breaks the inversion symmetry of the phosphorene, while the mechanical tension stress does not.

We now obtained that the large  enhancement of the PGE photocurrent  under the appropriate mechanical strain is primarily due to the substantially increased device asymmetry, while the change in the optical absorption has a weaker contribution to the enhancement in most of the cases. This point is in agreement with the recent experimental findings, which showed that the change in the asymmetry of the LiNbO$_3$ single-crystal by the mechanical compressive stress has a more important contribution on the enhancement of the PGE photocurrent than the change in the optical absorption.\cite{sci.adv.2019.5.eaau9199} Our results show that under an appropriate mechanical stress  the PGE photocurrent can be substantially enhanced even by up to 3 orders of magnitude. In this regard, it is understandable that in the flexo-photovoltaic effect the mechanical gradient can induce a considerable enhancement of the PGE photocurrent for the SrTiO$_3$ single crystals.\cite{Science.2018.360} Moreover, in our work the mechanical bending is even more effectively in improving the PGE photocurrent, which is in a good agreement with the essential point derived from the recent experiments that reducing the structure symmetry can largely enhance the PGE photocurrent in the WSe$_2$ nanotubes.\cite{Nat.2019} Interestingly, recent experiments have also shown that the photocurrent generated in the flexible  2D PtSe$_2$ phototransistor\cite{Small.2018.14} and the 2D antimonene photodetector\cite{Nano.Horiz.2019} can be enhanced by the mechanical bending, which suggested that the mechanical tuned PGE may play a role. Therefore, our results on the enhancement of the PGE by the mechanical strain can provide insights on these related experiments, though further theoretical efforts are required to give the direct supports.

\subsection{CONCLUSIONS}

In summary, we have investigated the large enhancement of the PGE in the nickel-phosphorene-nickel photodetector tuned by the mechanical tension and bending, using  the quantum transport calculations. The photocurrent  can be substantially improved by 3 orders of magnitude by applying the appropriate  mechanical tension  and bending. The remarkable enhancement of the photocurrent is overall mainly due to the largely increased device asymmetry, although the change in the optical absorption can also play an important role in some cases.  The mechanical bending can enhance the photocurrent more effectively than the tension stress, since it breaks the inversion symmetry of the phosphorene and hence strengthens largely the device asymmetry. Our results propose an approach to largely enhance the PGE photocurrent by applying the mechanical tension and bending, and shed lights on the applications of the PGE in  the low-power 2D flexible optoelectronics and low-dimensional photodetections with high photoresponsivity, as well as give insights to the enhancement of the PGE achieved recently in experiments on the nanotubes and  flexo-photovoltaic effects.

\section*{\textbf{ACKNOWLEDGMENT}}
This work is supported by National Natural Science Foundation of China under Grant No.51871156 and 11704232, and also by National Key R\&D Program of China under Grants No. 2017YFA0304203, 1331KSC, Shanxi Province 100-Plan Talent Program.


\end{document}